\documentclass{article}

\usepackage{graphicx}
\usepackage{amsmath,amssymb}
\usepackage{algorithm}
\usepackage{algorithmic}

\begin{document}

\title{A Multiagent Simulation for Traffic Flow Management with~Evolutionary
Optimization}

\author{Patryk Filipiak\\Institute of Computer Science\\University of Wroclaw, Poland\\patryk.filipiak@ii.uni.wroc.pl}

\maketitle

\begin{abstract}
A traffic flow is one of the main transportation issues in nowadays
industrialized agglomerations. Configuration of traffic lights is
among the key aspects in traffic flow management. This paper proposes an
evolutionary optimization tool that utilizes multiagent simulator in order to
obtain accurate model. Even though more detailed studies are still necessary, a
preliminary research gives an expectation for promising results.
\end{abstract}

\section{Traffic flow management}

Early models of traffic flow (called
\emph{macroscope} models) treated vehicles in the collective manner basing on
the analogy to particles in a fluid \cite{lighthill}. Later on, more precise
(\emph{mezoscope}) models were being created consecutively (mostly based on gas
kinetics) \cite{state_of_the_art}. Currently, a multiagent simulation mechanism
provides much more efficient \emph{microscope} model where each vehicle can be
regarded separately allowing for highly detailed analysis including collision
avoidance \cite{collision}, traffic virtualization \cite{virtual}, interactions
with pedestrians \cite{pedestrians}, etc.

Traffic flow management varies from tracing main roads average capacity across
certain area (where macroscope models give satisfactory results) to very
low-level manipulations including re-arrangement of lanes, modifying traffic
lights configuration, planning bridge locations, etc. where microscope models
are most suitable \cite{sroka}.

\section{Multiagent traffic flow simulator}

\begin{figure}[t!]
\begin{center}
\includegraphics[width=8cm]{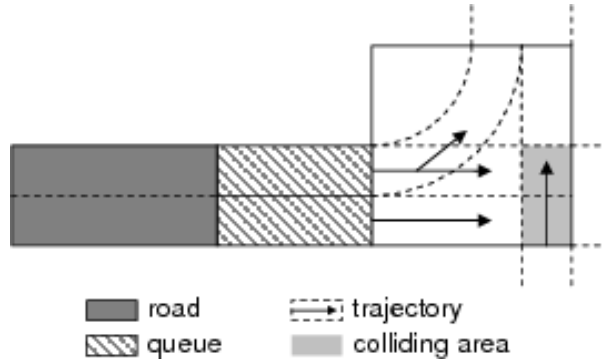}
\caption{Elements of environment description presented on the part of a
sample junction.}
\label{figure:representation}
\end{center}
\end{figure}

The agent-based traffic flow simulator described in \cite{sroka} is the
universal microscope model. The environment for agents in this model comprises
of a system of city streets defined in XML files using the following entities
illustrated in Figure~\ref{figure:representation}:

\begin{itemize}
  \item \textbf{road} -- defines the one-way part of a street, i.e. an existing
  path leading from one junction to another. For the sake of simplicity, roads
  at the frontiers of a model are considered dead-end even though they play
  roles of ingoing or outgoing roads.
  \item \textbf{queue} -- defines a single lane represented as a
  first-in-first-out list of vehicles located on this lane one after another.
  Note that there has to be at least one queue for each road at each junction. 
  \item \textbf{trajectory} -- corresponds to a trajectory of vehicles crossing
  the junction from a given ingoing road to a given outgoing road. Each
  trajectory stores the information about other trajectories with possibile
  collisions with it.
  \item \textbf{track} -- an abstract pair \emph{queue-trajectory} binding three
  entities, namely queue at ingoing road, queue at outgoing road and trajectory
  from one to another.
\end{itemize}

Each vehicle in the environment is an agent that is aware of its current speed
and exact location within a model (particularly, its relative position on the
present road). What is more, a vehicle is assigned another agent who drives it
in order to achive the driver's own aim which is to get from point A to point B
chosing the best possible way. 

Driver agents are equipped with a simple perception mechanism that allows them
to acquire information about a vehicle driven by them (e.g. current lane,
driving speed or acceleration) and the current state of environment in their
proximity, particularly: other vehicles (their behaviour, distance to them,
etc.) and traffic lights. 

For the sake of efficiency, described model includes the following
simplifications that should be taken into account: there are traffic lights at
each junction, there are no roundabouts and the presence of pedestrians is
ignored.

Simulation process in described model is the sequence of $N \in \mathbb{N}$
iterations executed every $\tau > 0$ miliseconds (by default $N = 1000$ and
$\tau = 200$). Each iteration includes the following operations:
\begin{enumerate}
  \item Apply changes in the environment (if there are any at this time period)
  including the presence of colliding tracks and the state of traffic lights.
  \item Refresh all agents' local information about the state of environment.
  \item Refresh locations of all moving vehicles according to intentions of
  their drivers considering current speed and acceleration of vehicles.
  \item Remove agents that achived their aims and create new ones if possible.
\end{enumerate}

Vehicle agents collect some essential information during simulation, i.e. exact
trace; number, locations and durations of stops; average speed etc. for the sake
of further analysis.

\begin{figure}[t!]
\begin{center}
\includegraphics[width=8cm]{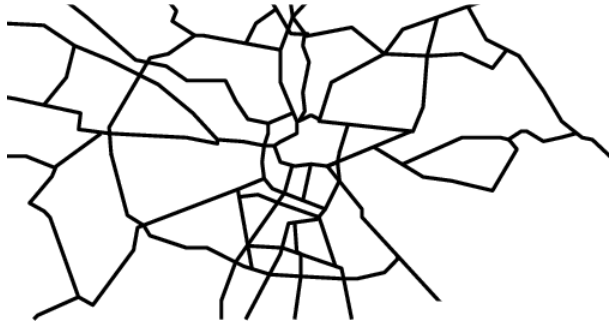}
\end{center}
\caption{Arrangement of main streets in Wroclaw (Poland) that can be easily
modelled with the agent-based simulator.}
\label{figure:wroclaw}
\end{figure}

Figure~\ref{figure:wroclaw} depicts arrangement of main streets in
Wroclaw, Poland that can be easily modelled using described agent-based
simulator.

\section{Optimization of traffic lights}

Traffic lights are commonly used as an effective solution for car flow routine
regulation in virtually all agglomerations. Modification of traffic lights
configuration is the key aspect in optimization of traffic flow for the model
presented in the previous section.

In many real-world situations, traffic lights configuration settings are
based on combination of averaged theoretical data and designer's assumptions. A
multiagent simulation approach can be used instead in order to provide more
accurate data. 

\begin{figure}[b!]
\begin{center}
\begin{tabular}{cc}
\includegraphics[width=2.5cm]{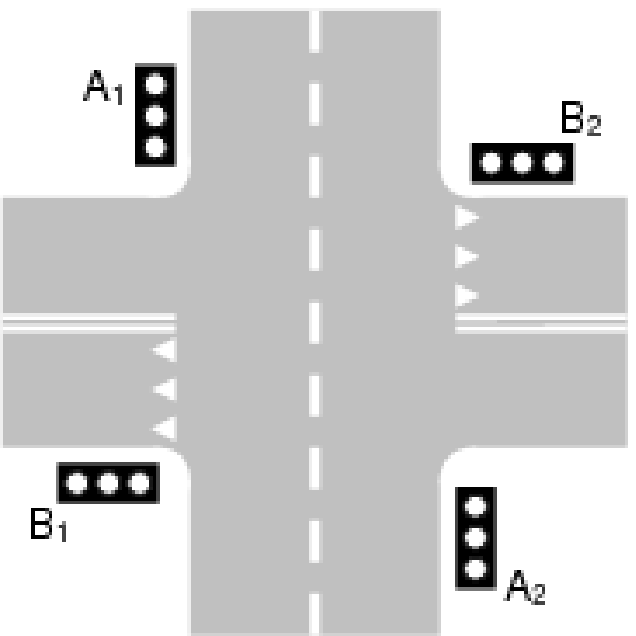}
&
\includegraphics[width=5.5cm]{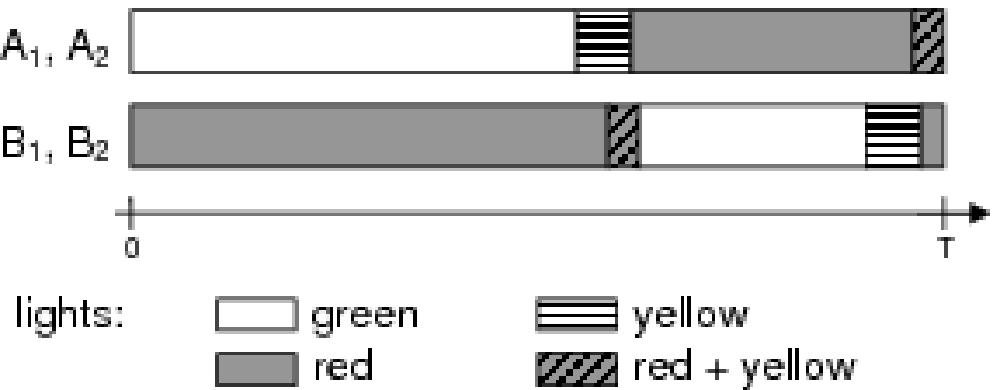}
\end{tabular}
\caption{A sample junction with two pairs of traffic lights
(left) and a corresponding traffic lights programme (right).}
\label{figure:programme}
\end{center}
\end{figure}

A cyclic sequence of consecutive changes of lights including detailed
information about duration of time when each light is on combined with
dependencies amongst all traffic lights within a given area (typically a
crossroad) is referred to as \emph{traffic lights programme}. The usual
representation of such programme is a graphical timeline presented in
Figure~\ref{figure:programme}.
 
The merit of underlying optimization process lies in realistic accuracy provided
by the agent-based simulator. Performance of contemporary hardware is sufficiently
enough for launching simulations containing even millions of agents in place of
less accurate statistical estimations that were used in previous years. Thus,
the values of a function that is being optimized are in fact results of
agent-based simulation.

\subsection{Constraints}

Any traffic lights programme is obliged to satisfy the following constraints:
\begin{itemize}
  \item A change of lights cannot violate the sequence (green, yellow, red,
red+yellow) repeated continuously.
  \item Duration of green light must be greater or equal to a given minimal time
  length whereas durations of transient lights (i.e. yellow and red+yellow) are
  constant and cannot undergo any optimization.
  \item No collision possibilities are allowed, i.e. red light must be lit on at
  one pair of traffic lights until green light turns to yellow on another pair and remains yellow for a given time period; similar condition must be
satisfied during a change from green to red light.  
\end{itemize}

\subsection{Criteria}

The aim is to optimize a traffic lights programme by maximizing a car flow
subject to constraints described above. Typical optimization criteria in such
case would be: maximizing the number of agents that achived their aims during
simulation, minimizing the average time of a ride, maximizing average speed of
vehicles, etc. Since most of the criteria correspond to each other an
aggregation of at least two of them provides a fair evaluation function.

\begin{figure}[t!]
\begin{center}
\includegraphics[width=8cm]{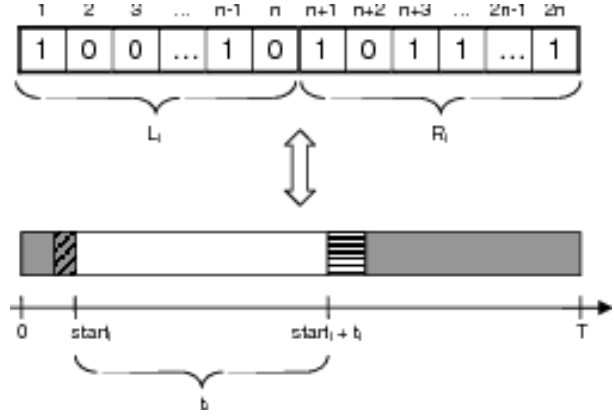}
\caption{A sample programme of $i$-th traffic lights represented in binary
chromosome form (up) and its corresponding timeline form (down).}
\label{figure:coding}
\end{center}
\end{figure}

\subsection{Representation}

Assume that $T > 0$ is a time length of one traffic lights cycle at a junction.
Time range $[0, T]$ can be discretized into $T / \tau$ intervals of $\tau > 0$
miliseconds each as it was mentioned in the previous section. Let $t_{min}$ be
the minimal and
\begin{equation}
t_{max} = \frac{T}{\tau} - k \cdot t_{min} - C, \quad (C = \mbox{const})
\end{equation}
be the maximal feasible time length of green light duration. Note that $t_{max}$
is determined by the time length of green light duration for $k$ tracks
colliding with current track and some constant value reserved for transient lights.

As a result, any programme of $M > 0$ traffic lights can be well-defined by the
set of pairs
\begin{equation}
\{ (start_1, t_1), (start_2, t_2), \ldots, (start_M, t_M) \},
\label{equation:programme_representation}
\end{equation} 
where $start_i$ represents the number of time interval when $i$-th traffic light
turns to green and remains for the next $t_i$ time intervals ($i = 1, \ldots,
M$). Duration of red light can be computed therefore as $T - t_i - C$.

Evolutionary algorithms provide an efficient optimization mechanism for the
problem presented above. Let $n > 0$ be the lowest integer value that satisfies
condition $T / \tau \leq 2^n$, hence
\begin{equation}
0 < t_{min} \leq t_{max} \leq 2^n.
\end{equation}

Any traffic lights programme expressed in the form
(\ref{equation:programme_representation}) can be encoded as a $2MN$-element
binary chromosome containing $M$ pairs of $N$-bit values $(L_i, R_i)$ such that
\begin{eqnarray}
start_i &=& L_i \mod T/\tau, \\
t_i &=& t_{min} + [R_i \mod (t_{max} - t_{min} + 1)] 
\end{eqnarray} 
for $i = 1, \ldots, M$.

\subsection{Algorithm}

Population-Based Incremental Learning (PBIL) \cite{pbil} can be used as an
evolutionary optimization tool for the stated problem thanks to its simplicity
and promising results in a broad range of applications. It is also worth
noticing that even more complex evolutionary algorithms could be applied instead without a significant decrease of performance due to
considerably time consuming evaluation process that is suggested in this paper. 

Algorithm~\ref{algorithm:pbil} presents the pseudocode of PBIL. It is so called
Estimation of Distribution Algorithm (EDA) \cite{eda} based on the idea of
optimizing a probability model for solution rather than implicit set of
solutions. Later on, individuals are randomly generated according to obtained
probability distribution.

\begin{figure}[b!]
\begin{center}
\includegraphics[width=8cm]{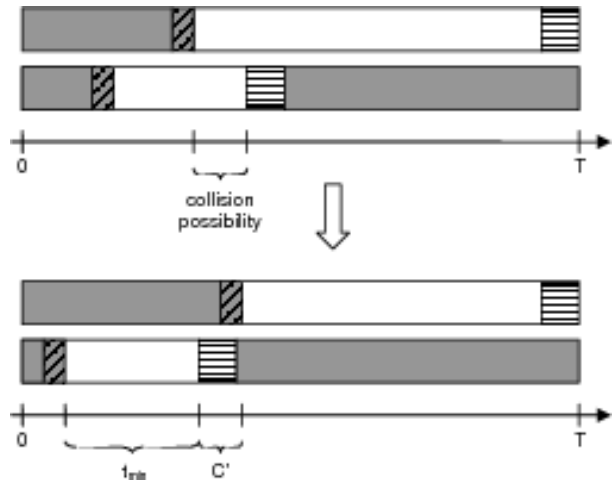}
\caption{Simple conflicts resolving method.}
\label{figure:conflicts}
\end{center}
\end{figure}

The requirement for collision avoidance is not satisfied by the traffic lights
programme definition itself, hence the presence of infeasible individuals in
population might be expected. Figure~\ref{figure:conflicts} presents a simple
conflicts resolving method for this case. If a collision possibility occurres
between a pair of trajectories then corresponding traffic lights programme can
be modified in order to ensure arbitrary set $C'$ distance between green light
durations. If any of modified green light durations becomes shorter than
$t_{min}$ then it is automaticaly expanded to $t_{min}$ in the only possible
way. Finally, if a conflict still occurres, a new random individual is
selected.

\begin{algorithm}
\caption{Standard PBIL algorithm with parameters $0 < \theta_1, \theta_2,
\theta_3 < 1$ on a population of $size$ individuals each of which is
represented as $d$-element binary vector.}
\label{algorithm:pbil}
\begin{algorithmic}
\STATE $p \leftarrow$ InitialProbabilityVector()
\STATE $P \leftarrow$ RandomPopulation($p$, $size$) 
\STATE PopulationEvaluation($P$)
\WHILE {\NOT TerminationCondition($P$)} 
	\STATE $x_i \leftarrow$ BestIndividual($P$)
	\FOR {$k \leftarrow 1$ \TO $d$}
		\STATE $p_k \leftarrow p_k \cdot (1 - \theta_1) + x_{ik} \cdot \theta_1$
	\ENDFOR
	\FOR {$k \leftarrow 1$ \TO $d$}
		\IF {UniformRandom(0, 1) < $\theta_2$}
			\STATE $p_k \leftarrow p_k \cdot (1 - \theta_3) +
			\mbox{BinaryRandom(0.5)} \cdot \theta_3$
		\ENDIF
	\ENDFOR
	\STATE $P \leftarrow$ RandomPopulation($p$, $size$) 
	\STATE PopulationEvaluation($P$)
\ENDWHILE
\end{algorithmic}
\end{algorithm}

\section{Conclusions and Perspectives}

Even though the results are partial at the moment and more detailed studies are
still necessary, a preliminary research gives an expectation for promising
results. An improvement of agent-based traffic flow model could be a next step. Also a modification of evolutionary
mechanism may improve the performance of optimization process. Replacing PBIL
with Bayesian Optimization Algorithm (BOA) \cite{boa} could ensure obtaining
much more detailed model. Unlike other evolutionary algorithms, BOA not
only searches for optimal solution but also gives some information about its
structure. Lots of dependencies (e.g. among traffic lanes and even adjacent
junctions) could be modelled in terms of Bayesian networks that are utilized in BOA.
What is more, such dependency model could include the fact that more effort
should be put to traffic flow management inside city centers than suburbs
because of greater amount of junctions, narrow and/or one-way street, etc. 

As it is clearly seen, the effectiveness of agent-based approach applied for
traffic flow managements provides a broad range of ideas.

\section{Acknowledgements}

The multiagent simulator developed by Krzysztof Sroka within confines of his
Master's dissertation \cite{sroka} was extensively used during research process
for this paper.

\end{document}